\documentclass{IEEEtran4PSCC}

\newcommand\scalemath[2]{\scalebox{#1}{\mbox{\ensuremath{\displaystyle #2}}}}
\usepackage{xcolor}
\ifCLASSINFOpdf
   \usepackage[pdftex]{graphicx}
\else
   \usepackage[dvips]{graphicx}
\fi
\usepackage[tight,footnotesize]{subfigure}
\usepackage[cmex10]{amsmath}
\usepackage{bm}
\usepackage{makecell}
\usepackage{multirow}
\makeatletter
\let\old@ps@headings\ps@headings
\let\old@ps@IEEEtitlepagestyle\ps@IEEEtitlepagestyle
\def\psccfooter#1{%
    \def\ps@headings{%
        \old@ps@headings%
        \def\@oddfoot{\strut\hfill#1\hfill\strut}%
        \def\@evenfoot{\strut\hfill#1\hfill\strut}%
    }%
    \def\ps@IEEEtitlepagestyle{%
        \old@ps@IEEEtitlepagestyle%
        \def\@oddfoot{\strut\hfill#1\hfill\strut}%
        \def\@evenfoot{\strut\hfill#1\hfill\strut}%
    }%
    \ps@headings%
}
\makeatother

\begin{document}
%
\title{Derisking of subsynchronous torsional oscillations in power systems with conventional and inverter-based generation}

\author{
\IEEEauthorblockN{Nicolas Bonafé, Julian Freytes}
\IEEEauthorblockA{EDF R\&D \\
Palaiseau, France}
\and
\IEEEauthorblockN{Hani Saad}
\IEEEauthorblockA{ACDC Transient \\
Lyon, France}
}


\maketitle

\begin{abstract}
This article proposes an application of a derisking methodology of subsynchronous torsional oscillations considering a realistic use case. The main objective is to summarize and draft a synthetic paper clarifying the complete methodology highlighting the main information needed step-by-step. For exemplification, a real model from a decommissioned oil power plant is adopted, where a fictitious high voltage direct current power link is connected. In this article, stress is laid on details of the application of the derisking methods: the unit interaction factor and the complex torque coefficients method. Then, the different steps to obtain results are explicitly explained. Moreover, the design and tuning process of supplementary subsynchronous damping controller is discussed. This mitigation section uses minimal information to correctly damp the unstable oscillations, as one would expect from industrial projects where the data sharing may be limited. Finally, the resources needed to perform each step of the study were summarized.

\end{abstract}



\section{Introduction}
The subsynchronous oscillations (SSO) in general, and especially the subsynchronous torsional interaction (SSTI), has been a major issue for decades \cite{cheng_real-world_2023}. SSO can be caused by the interaction of different systems of the electrical network \cite{damas_subsynchronous_2020}. On the one hand, one of the most frequent is the interaction between mechanical shafts and series compensated lines. On the other hand, SSTI, are caused by interactions between the mechanical shaft and the control of power electronics (HVDC, wind farm, etc.) at the subsynchronous modal frequencies of the shaft. This is the main phenomena of interest in this paper.
One of the first SSTI reported event was at the LCC HVDC station Square Butte in 1977 \cite{kovacevic_analysis_2020}. Back then, the HVDC link power control was found responsible of the interaction with the power plant. Later, there have been several other SSTI events between LCC-HVDC and generator worldwide. SSTI can induce loses of production or even degradation of the installations. The development of large HVDC links and Inverter-Based Generation (IBR) near the existent conventional generation obliges to assess and derisk any possible SSTI risks; particularly when equipments are electrically close and of comparable electrical sizes. That is why it is important to revisit methodology to prevent and to mitigate the risks of interaction between old and new inverter-based installations.

Different methods exists to evaluate risk of SSO \cite{damas_subsynchronous_2020, c_karawita_c4-b4_2023}. Still, not all methods are readily usable for this type of studies for varied reasons. The chosen methods must be accurate and robust, but also suitable from an industrial point of view since several data may be missing. Even though methods like eigen-based analysis are interesting to study SSO stability, they require detailed information of the physical plant model and their control algorithms. This type of information is not always available and are usually protected by Non-Disclosure Agreement (NDA), hence, not exploitable for the classical analytical methods \cite{Kovacevic2022}. Thus, it is important to consider methods which are readily usable for industrial studies, and to list the resources needed to perform the studies. The models used in a SSO study are supposed black box but with a quality allowing the SSO risk studies.
The selected methods will be based on Electromagnetic Transient (EMT) simulations.

The contribution of this article is the applicability of a SSTI derisking methodology to a detailed industrial based use case, focusing on the needed information for each step. Moreover, mitigation methods of SSTI instabilities through hardware modifications and control tuning methods are presented. The implementation and tuning method of a damping control in the converter is discussed. Finally, this article highlights the information needed to apply the presented method: data, computing resources and interactions between system stakeholders.

\section{Test case definition for exemplification of derisking method}

\subsection{Aramon power plant with an HVDC link electrically close}

\begin{figure}[!htbp]
\centering \includegraphics[width=0.8\columnwidth]{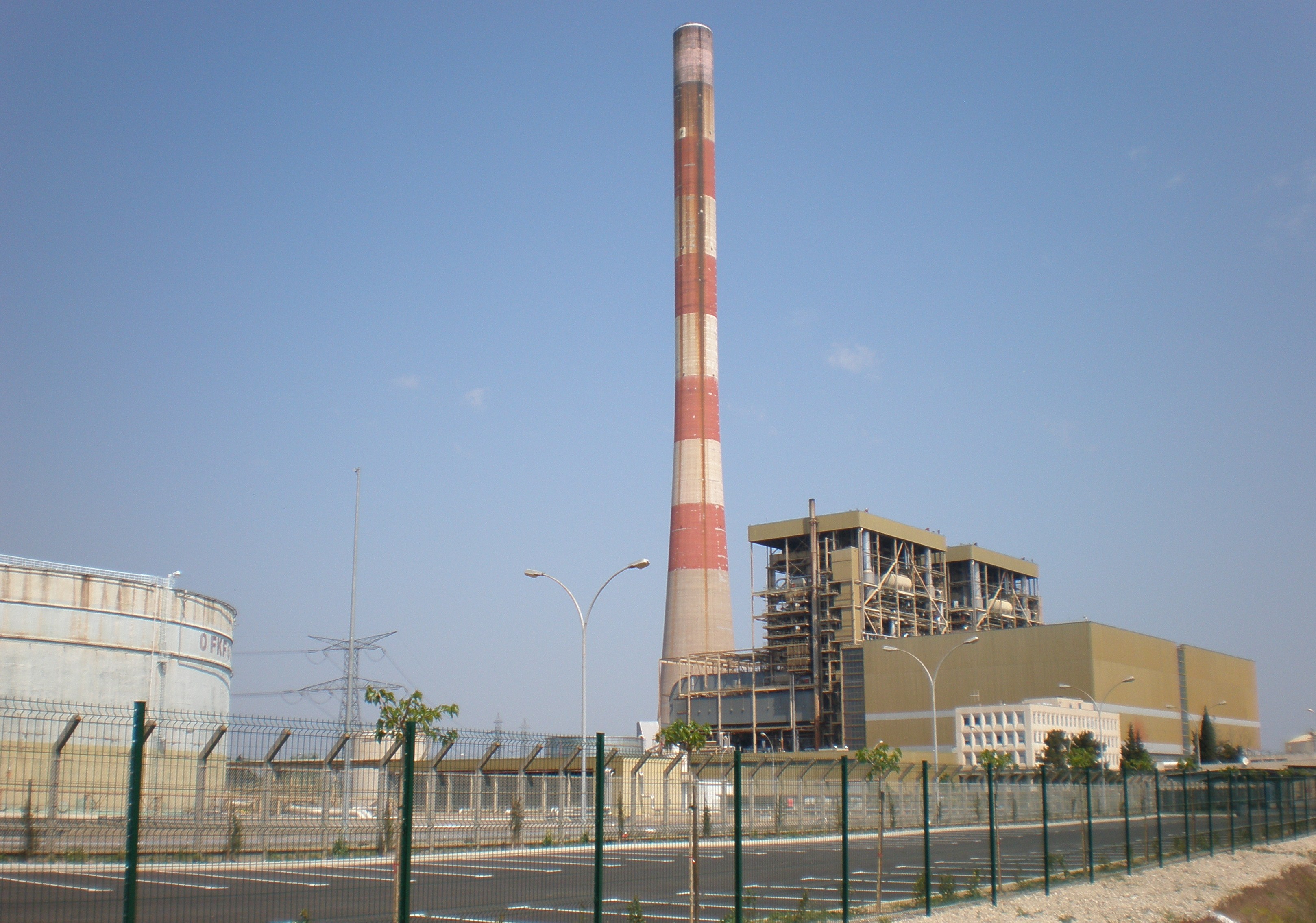}
\caption{Aramon power plant.}
\label{fig:Aramon}
\end{figure}

\begin{figure}[!htbp]
\centering
\includegraphics[width=0.99\columnwidth]{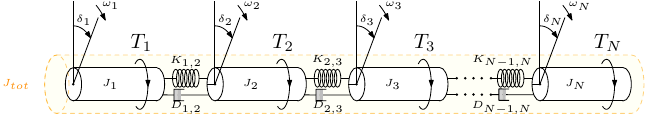}
\caption{Multi-mass model.}
\label{fig:Lumped_mass}
\end{figure}

\begin{figure}[!htbp]
\centering
\includegraphics[width=0.99\columnwidth]{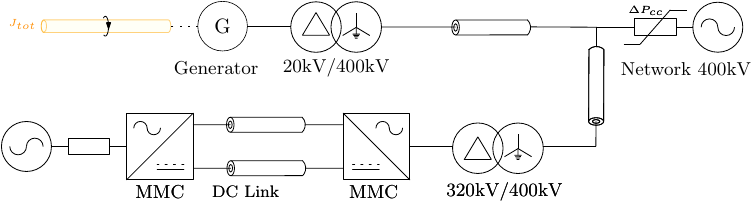}
\caption{Study case considering Aramon generator and HVDC link.}
\label{fig:study_case}
\end{figure}

To emphasize the methodology process and resource needs, a realistic use case will be considered. The Aramon power plant (see Fig.~\ref{fig:Aramon}) was a French $2\times778$~MVA oil plant decommissioned in 2016. Electrical and mechanical data of this power plant are used to model the system in EMTP-RV (see Tab~\ref{table_data}). The considered full scheme also includes a 1~GW MMC-HVDC link connected electrically close to the power plant, as depicted in Fig.~\ref{fig:study_case}. The HVDC link is modelled with the generic model provided in EMTP-RV library \cite{Saad2014}. Only one 778~MVA machine will be used. Both systems are connected to a 400~kV network. Moreover, the mechanical model of the generator includes the option to consider a lumped single mass representation of the rotor or a multi-mass model which allows SSO investigations (see Fig.~\ref{fig:Lumped_mass}).

\begin{figure}[!htbp]
\centering
\includegraphics[width=0.99\columnwidth]{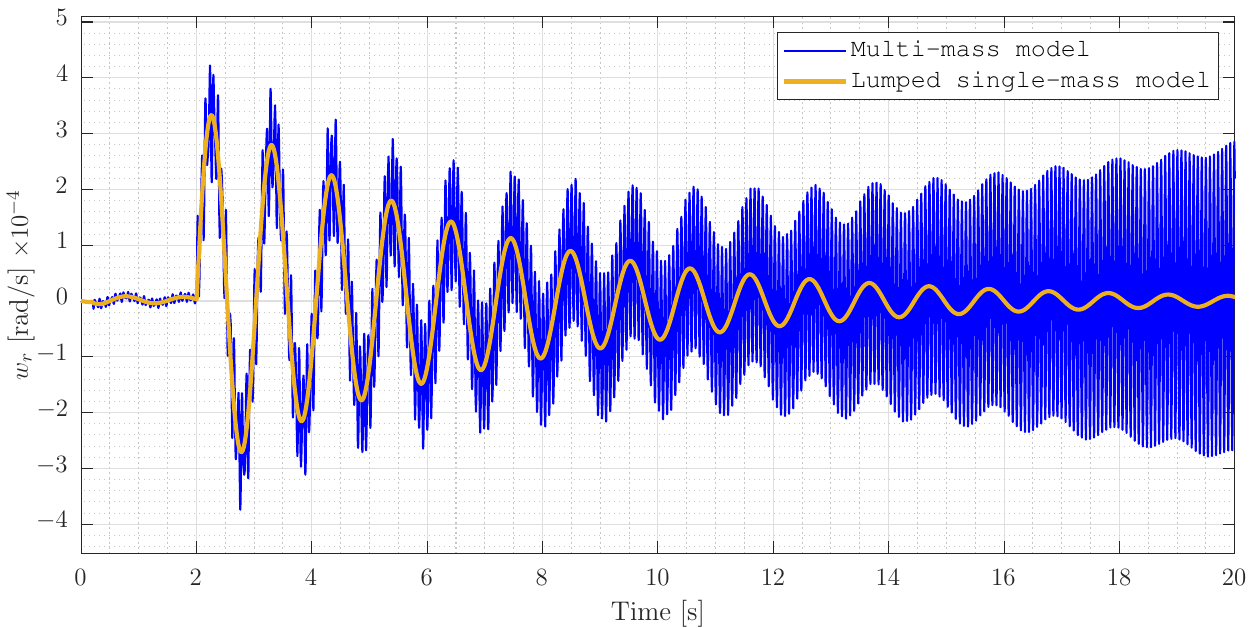}
\caption{Rotor speed deviation comparison with single and multi-mass model.}
\label{fig:wr_without_SSDC}
\end{figure}

\begin{table}[!htbp]
\renewcommand{\arraystretch}{1.3}
\setlength{\tabcolsep}{1pt}
\centering
\caption{Study case data.}

\label{table_data}
\begin{tabular}{|c|c|c|c|c|c|c|}
\hline
Names & \multicolumn{2}{c||}{Values} & \multicolumn{2}{c|}{Names} & \multicolumn{2}{c|}{Values}\\
\hline
Ligne impedance & \multicolumn{2}{c||}{$0.27\Omega/\text{km}$} & \multicolumn{2}{c|}{$\Delta S_{sc}$ step} & \multicolumn{2}{c|}{$-200\text{~MVA}$}\\
\hline
\makecell{Aramon-Network \\ distance} & \multicolumn{2}{c||}{$10\text{km}$} & \multicolumn{2}{c|}{$S_{Aramon}$} & \multicolumn{2}{c|}{$778\text{~MVA}$}\\
\hline
\makecell{HVDC-Network \\ distance} & \multicolumn{2}{c||}{$30\text{km}$} & \multicolumn{2}{c|}{$S_{\text{HVDC}}$} & \multicolumn{2}{c|}{$1000\text{~MVA}$}\\
\hline
\makecell{Network initial \\ $S_{sc}$} & \multicolumn{2}{c||}{$1550\text{~MVA}$} & \multicolumn{2}{c|}{\makecell{HVDC link \\ length}} & \multicolumn{2}{c|}{$70\text{km}$}\\
\hline
\makecell{Moment of \\ inertia} & $J_1$ & $J_2$ & $J_3$ & $J_4$ & $J_5$ & $J_6$ \\
\hline	
Values (kg/$m^2$)& $1293$ & $4321$ & $22249$ & $22249$ & $10402$ & $176$\\
\hline
Mutual stiffness & $K_{12}$ & $K_{23}$ & $K_{34}$ & $K_{45}$ & $K_{56}$ & \multirow{4}{*}{ }\\
\hline	
\makecell{Values \\ (N/m $\times 10^8$)} & $1.134$ & $2.478$ & $1.653$ & $1.033$ & $0.071$ & \\
\hline
\makecell{Mutual \\ damping} & $D_{12}$ & $D_{23}$ & $D_{34}$ & $D_{45}$ & $D_{56}$ &\\
\hline	
Values (N.s/m)& $5675.0$ & $12395.0$ & $8265.0$ & $5165.0$ & $355.0$ & \\
\hline
Modes & $f_1$ & $f_2$ & $f_3$ & $f_4$ & $f_5$ &\\
\hline	
Frequency (Hz) & $14.07$ & $22.092$ & $32.341$ & $34.933$ & $58.772$ &\\
\hline
Mechanical damping & $D_{m1}$ & $D_{m2}$ & $D_{m3}$ & $D_{m4}$ & $D_{m5}$ &\\
\hline
Values (pu) & $0.98$ & $4.74$ & $86.94$ & $7140$ & $3.72 \times 10^7$ &\\
\hline
\end{tabular}
\end{table}
To highlight the need of considering multi-mass model for torsional interactions, the system from Fig.~\ref{fig:study_case} is simulated using the parameters from Table~\ref{table_data}, where, at $t=2s$ a change in the network induces a reduction of the short circuit power $\Delta S_{sc}$ of $-200\text{~MVA}$. Results are shown in Fig.~\ref{fig:wr_without_SSDC}. With the lumped single mass model (yellow part in Fig.~\ref{fig:Lumped_mass}), the system is stable. On the contrary, with the multi-mass model which takes into account torsional deformations, the risk of SSO can be asserted, as shown by the clear instability in the simulated results. The explanation of such SSTI is presented
in Section \ref{Unstability}. This comparison highlights the frequency of oscillation of the torsional modes and  explains the main reason why this happens.

\subsection{Unit Interaction Factor}

The Unit Interaction Factor (UIF) is a coefficient quantifying the interaction likelihood between systems, which was introduced in 1982 by the EPRI \cite{UIF}. This coefficient is used in pre-study stage to perform a network screening in order to identify the risk of interaction between power electronic device and power plant. UIF is a well know indicator because it is easy to use and straightforward to compute. It takes into account two factors: the size of the installation and the electrical distance. For a generator radially connected to the converter bus \cite{c_karawita_c4-b4_2023}: 
\begin{align}
 \text{UIF}_i = \frac{S_{\text{HVDC}}}{S_{\text{Gen}}} \left(1-\frac{S_{sc-i}}{S_{sc}}\right)^2\label{eq:UIF}
\end{align}

\noindent where $S_{HVDC}$ and $S_{Gen}$ (in MVA) are the nominal power of the installations. $S_{sc-i}$ ($sc-i$ means "minus" $i^{th}$) is the short circuit power without the $i^{th}$ machine connected and $S_{sc}$ is the total short circuit power at the commutating HVDC bus. The interaction risk is expected to be high (not necessary harmful) when the value of the $\text{UIF}_i$ is large; the common empiric threshold value for Line Commutated Converter (LCC) is 0.1. If $\text{UIF}_i$ is above $0.1$, it indicates that strong interactions are likely; further studies are needed to determine the nature of the interactions and the associated risks. In the other case, interactions are weak, and the risk remains low. However, for VSC-HVDC this threshold value is maybe inappropriate as highlighted in \cite{c_karawita_c4-b4_2023}. Then, the results obtained by UIF with a VSC are not very conclusive.

For the Aramon use case, the calculated value is $\text{UIF} = 0.44 > 0.1$. According to the literature, the risk is high, hence a detailed SSO risk study must be performed. The UIF is not taking into account control loops and detailed models. Therefore, further studies are needed.

\subsection{Complex torque coefficients method} 

\begin{figure}[!htbp]
\centering
\includegraphics[width=0.6\columnwidth]{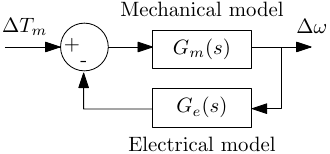}
\caption{Interaction model \cite{harnefors_analysis_2007}.}
\label{fig:Interaction_model}
\end{figure}

In case the UIF screening study highlights a strong interaction risk, a more precise study is applied (e.g using EMT simulations). In this article, the complex torque coefficients method is used to evaluate the stability of the system, and was developed in 1982 in \cite{Canay1982} \cite{Canay1982a2}. The objective is to determine the stability of the system through the measured damping of the system. The simplified structure Fig.~\ref{fig:Interaction_model} is used to model the interaction between the electric network and the mechanical part \cite{harnefors_analysis_2007}. This model remains limited as it does not take into account non-linearities but is enough for this type of studies. The expression of the transfer functions is:
\begin{align}
 G_e(s) = \frac{\Delta T_e}{\Delta \omega}= D_e(s) + jK_e(s)\label{eq:Ge}
\end{align}
\begin{align}
 \frac{1}{G_m(s)} = D_m(s) + jK_m(s)\label{eq:Gm}
\end{align}

\noindent where $D_e$ is the electrical damping and $D_m$ is the mechanical damping. They represent respectively the grid and the mechanical shaft participation to the damping of the SSO. The criterion to ensure a stable situation is:
\begin{align}
D_t = D_e (j\omega)+D_m (j\omega)>0\label{eq:Crit}
\end{align}
The total damping positiveness, $D_t$, needs to be verified at resonance frequencies and for low frequencies \cite{harnefors_analysis_2007}.

This method requires the computation of three elements: the mechanical damping, the resonance frequencies and the electrical damping.

\subsubsection{Mechanical damping calculation}

The mechanical damping is a pure mechanical coefficient which depends only on the mechanical multi-mass shaft model of the synchronous machine (which also considers the mechanical damping from the turbines and other elements in the shaft). As previously introduced in previous section, a multi-mass model of the rotor power plant is used to calculate the mechanical damping but also the resonance frequencies. Considering the model from Fig.~\ref{fig:Lumped_mass} of multi-mass model and applying $2^{nd}$ Newton's Law, the following system of equations can be obtained:
\begin{align}
 \bm{J}\ddot{\delta} + \bm{D}\dot{\delta}+\bm{K}\delta = \bm{T_m} - \bm{T_e}\label{eq:PFD_lumped_mass}
\end{align}
\noindent where the matrix of inertia moment: $\bm{J} = diag(J_1, J_2, ..., J_N)$, and $\bm{K}$ being the stiffness matrix and $\bm{D}$ the damping matrix defined as follows: 
\begin{align}
\scalemath{0.9}{
\bm{K} = \begin{pmatrix} 
K_{12} & -K_{12} & 0 & 0 & 0 \\ 
-K_{12} & K_{12} + K_{23} & -K_{23} & 0 & \vdots\\
0 & \ddots & \ddots & \ddots & 0\\
\vdots & \ddots & \ddots & \ddots & K_{N-1N}\\
0 & 0 & 0 & -K_{N-1N} & K_{N-1N}
\end{pmatrix}}
\end{align}
\begin{align} 
\scalemath{0.9}{
\bm{D} = \begin{pmatrix} 
D_{12} & -D_{12} & 0 & 0 & 0 \\ 
-D_{12} & D_{12} + D_{23} & -D_{23} & 0 & \vdots\\
0 & \ddots & \ddots & \ddots & 0\\
\vdots & \ddots & \ddots & \ddots & D_{N-1N}\\
0 & 0 & 0 & -D_{N-1N} & D_{N-1N}
\end{pmatrix}}
\end{align}
\noindent The expression from \eqref{eq:PFD_lumped_mass} can be express in a state equation expression as follows:
\begin{align}
 \dot{\bm{X}} = \bm{A}\bm{X}+\bm{B}\bm{T}\label{eq:State_rep}
\end{align}

\noindent with $\bm{X}=(\dot{\delta_1}, ..., \dot{\delta_N}, \delta_1, ..., \delta_N)^T$, and:

\begin{align}
\bm{A}=\begin{pmatrix} 
-\bm{J}^{-1}\bm{D} & -\bm{J}^{-1} \bm{K} \\ 
\bm{I_{N\times N}} & \bm{0_{N\times N}} 
\end{pmatrix}, ~~~
\bm{B} = \begin{pmatrix} 
-\bm{J}^{-1} \\ 
0
\end{pmatrix}
\end{align}

The modal frequencies at which the shaft will oscillate are given by:
 \begin{align}
 \omega_i = Im(eig(\bf{A})_i)\label{eq:fn}
\end{align}
\begin{align}
 \sigma_i = Re(eig(\bf{A})_i)\label{eq:sign}
\end{align}
To compute the mechanical damping, the modal inertia needs to be calculate first. It is obtained by doing a base transformation of the equation system using \cite{padiyar_analysis_1999}:
\begin{align}
 \bm{Q} = \text{eigenvectors}(\frac{1}{2} \bm{H}^{-1}\bm{K_{pu}})\label{eq:Q}
\end{align}
\noindent where $\bm{Q}$ is normalized for the coefficients corresponding to the generator. For example, if the generator is the $5^{th}$ mass of the lumped-mass model, the $5^{th}$ line of $\bm{Q}$ must be equal to 1. $\bm{K_{pu}}$ is the stiffness matrix in per unit. Then, to obtain $\bm{H_{m}}$ the model inertia, the following expression is used:
\begin{align}
 \bm{H_{m}} = (\bm{H}^\frac{1}{2} \bm{Q})^T(\bm{H}^\frac{1}{2} \bm{Q})\label{eq:Hmi}
\end{align}
\noindent with $\bm{H_{m}}=(H_{m1}, ..., H_{mN})^T$.
Finally, one can compute the mechanical damping $\bm{D_{m}}$ as follows \cite{Canay1982, damas_subsynchronous_2020}:
\begin{align}
 D_{mi} = -4\times \sigma_i\times H_{mi}\label{eq:Dmi}
\end{align}
This calculus is direct and doesn't require heavy computing as matrix have low dimensions. The mechanical damping values can be found Table~\ref{table_data}.

\subsubsection{Electrical damping $D_e$ computation methods}
The specificity of this method is the computation of the electrical damping which is based on transient simulations. One can imagine computing it explicitly via analytical calculations supposing that a full-order state-space model of the system can be built. However, in this article it is considered that one doesn't have access to the entire system model but rather the numerical models of the assets. So, time-domain simulation is a direct method to obtain the value of the electrical damping at different frequency using black-box models.

\begin{figure}[!htbp]
\centering
\includegraphics[width=0.6\columnwidth]{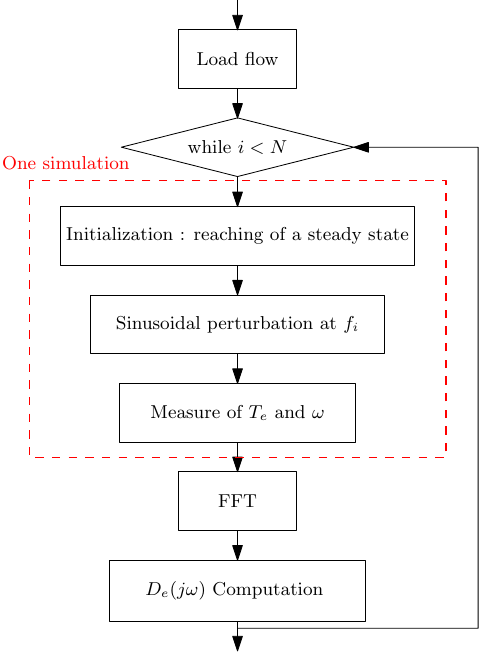}
\caption{Algorithm to compute the electrical damping.}
\label{fig:De_algo}
\end{figure}

As one can see in equation (\ref{eq:Ge}), the electrical damping can be obtained by computing the real part of the transfer function:
\begin{align}
 D_e(j\omega) = Re\left(\frac{\Delta T_e(j\omega)}{\Delta\omega(j\omega)}\right)=\frac{|\Delta T_e(j\omega)|}{|\Delta\omega(j\omega)|}cos(\phi_{Te} - \phi_\omega)\label{eq:De}
\end{align}
The objective is to extract the gain and the phase of the transfer function at different target frequencies. To do so, a small signal sinusoidal perturbation is injected in the mechanical torque at a chosen frequency. The electromagnetic torque and the rotation speed are then measured (after a fixed number of periods, for being sure that the transient is extinct and the steady-state is reached). The application of an FFT allows to recover the gains and phases to calculate the electrical damping $D_e$ at the perturbation frequency. The process is repeated iteratively as shown Fig.~\ref{fig:De_algo}.

This method uses a high number of simulations to obtain a high-fidelity electrical damping $D_e$ vs. frequency curve. At each step of the loop, a new simulation is started. And each time the same initialization process to reach steady state is computed. An advanced method would be to apply the same calculations but with a single simulation. The initialization process to steady state would be computed only once, hence accelerating the overall process. For the perturbations, instead of launching several simulations, only one is launched but with a progressive injection of the sinusoidal perturbation of different frequencies. The delay between each injection and the measurement must be long enough to not disturb the result. Moreover, to reduce the computing time, it is possible to inject several frequencies at the same time within the perturbation. Yet, the number of simultaneous frequencies is limited because of possible interactions between perturbations to avoid reaching non-linearities in the model.

The computation of a 50 points curve at a $20\mu s$ time step took 72 minutes. The use an advanced method could reduce this processing time. However, when using more complex models, the duration can be of several hours.

\subsection{Stability results}
\label{Unstability}
\begin{figure}[!htbp]
\centering
\includegraphics[width=0.99\columnwidth]{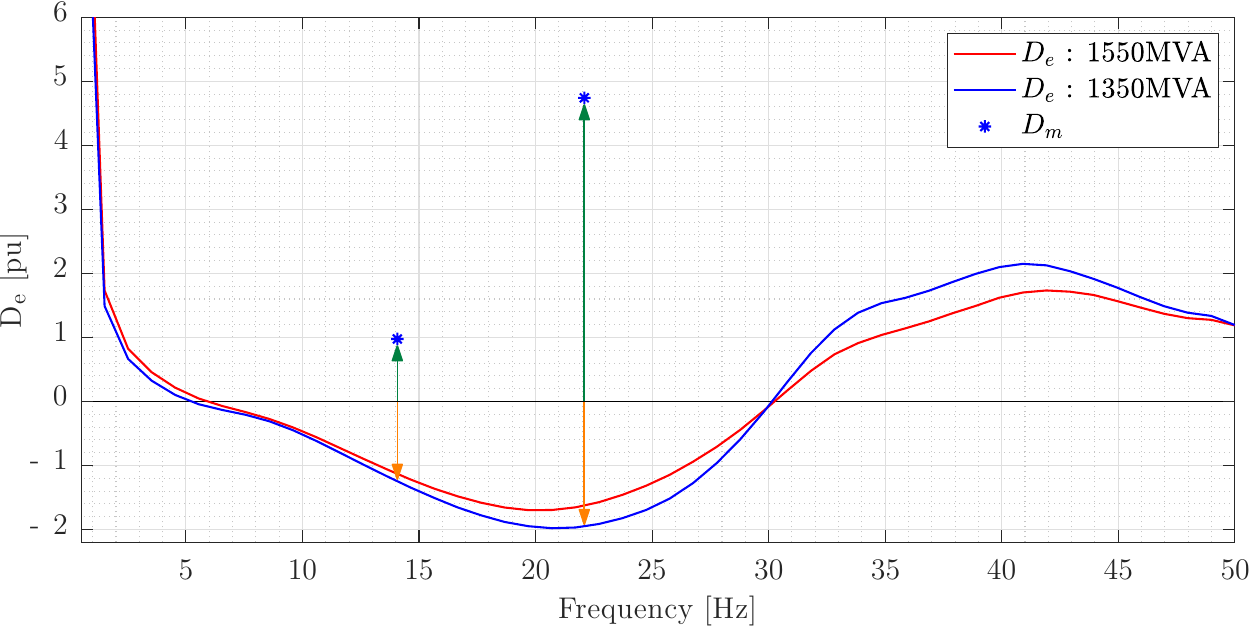}
\caption{Electrical damping in function of the frequency before (red) and after the network perturbation (blue).}
\label{fig:De_curve_init}
\end{figure}

Using this method the $D_e$ curve in function of the frequency is drawn Fig.~\ref{fig:De_curve_init} for the case of Aramon and the HVDC link. For the second and higher modes, the mechanical damping is large enough to validate the stability criterion given by (\ref{eq:Crit}). However, after the reduction of the short circuit power of the network, the first mode total damping becomes negative, i.e. $D_e + D_m = -0.2$, meaning that this mode is not stable. This explains the instability observed in Fig.~\ref{fig:wr_without_SSDC}.

\section{SSTI mitigation methods}

The result of the SSO risk study, using complex torque coefficients method, highlights the instability of the study case. The first mode is unstable and this can generate dangerous oscillations for the shaft of the power plant. Thus, solutions have to be found to mitigate the risk of undamped SSOs. In literature, many hardware and software-based techniques and methods exist \cite{wenpei_design_2020}. Hardware methods require the installation of new system such as STATCOM, SVC or static filters to damp the oscillations. Software-based methods consist in the design and tuning of control loop to improve the electrical damping at resonance frequencies.

\subsection{Blocking filters}

\begin{figure}[!htbp]
\centering
\includegraphics[width=0.6\columnwidth]{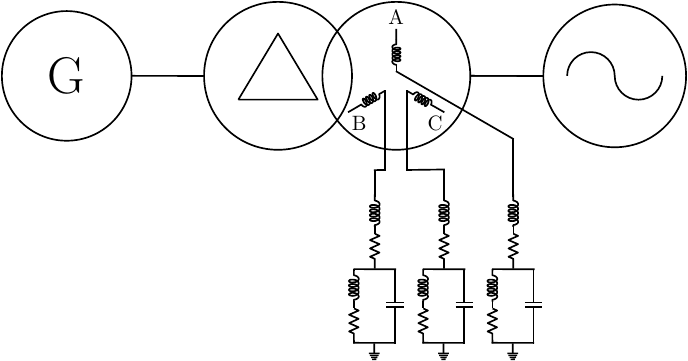}
\caption{Circuit diagram of static filters.}
\label{fig:Static_filters}
\end{figure}

The installation of additional passive filters to prevent interactions at the subsynchronous frequencies is a hardware solution to prevent SSO which can be applied on the power plant side. For example, the static blocking filters presented in \cite{c_karawita_c4-b4_2023} are added on the secondary side of the step-up transformer (Fig.~\ref{fig:Static_filters}). The chosen filters are notch filters. They have a high-quality factor and block a specific resonance frequency. A first limit to this mitigation method is the possible shift of the resonance frequencies due to external perturbations (temperature, network...) but also changes of the blocked frequency due to temperature sensitivity of the capacitors.

The blocking filter have a low impedance at system frequency and a high impedance at the blocked mode frequency. This allows an isolation of the generator at the problematic frequency.

\begin{figure}[!htbp]
\centering
\includegraphics[width=0.99\columnwidth]{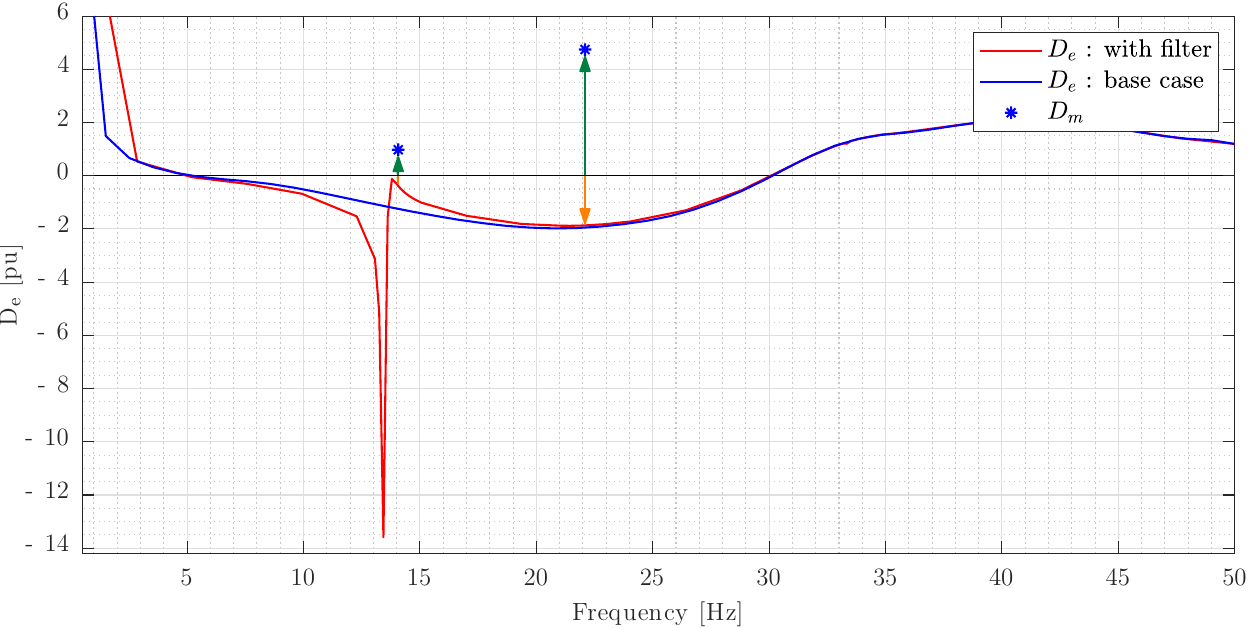}
\caption{Electrical damping curve with the blocking filters (red) and the base case without filters (blue).}
\label{fig:De_Filters}
\end{figure}

In the adopted study case in this paper, a blocking filter have been implemented in the step-up transformer. It mitigates the unstable mode at $14.07\text{Hz}$ (see Fig.~\ref{fig:De_Filters}). This solution requires the lowest uncertainty of the resonance frequency as possible. Indeed, the filter has a high-quality factor and is very selective. As a consequence, the $D_e$ value around $14.07\text{Hz}$ is unable to damp the SSO correctly. It is problematic if the resonance frequency changes too much.

\subsection{Supplementary subsynchronous damping controller (SSDC) design method}
\subsubsection{Definition}
A more versatile method to damp the SSO is to modify the control algorithms of the HVDC. The modification of the control loops to mitigate the SSO can be achieved by the tuning of parameters (such as PLL gains...) \cite{kovacevic_analysis_2020} but also by adding dedicated auxiliary controls. These controls, called Supplementary subsynchronous damping controller (SSDC), can be included in several systems of the grid: STATCOM, VSC, LCC... \cite{c_karawita_c4-b4_2023}. The main functioning of the SSDC is similar as the of Power System Stabilizer (PSS) but for higher frequencies (between some Hz up to tenths of Hz, depending on the desired frequencies to be damped). There are mainly two types of SSDC: the large-band which improves the damping of all modes and the narrow band which targets individual modes. By using narrow-band SSDC in parallel, it is possible to damp several modes at the same time. The obtained damping is reported to be more effective with the narrow-band option. To choose the input of this control any oscillating variable can be used: active power, rotor speed, network frequency. The output variable can be the active power, reactive power, AC voltage... 

\begin{figure}[!htbp]
\centering
\includegraphics[width=0.99\columnwidth]{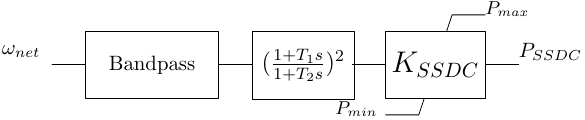}
\caption{Narrow-band SSDC, with input $\omega_{net}$ and output $P_{SSDC}$.}
\label{SSDC}
\end{figure}

This paper adopts the narrow-band SSDC, whose structure is depicted in Fig.~\ref{SSDC}. The band-pass filter isolates the harmonics to damp from the input signal. The lead-lag function shifts the phase to improve the damping. Then, the parameters to optimize are the band-pass parameters, the lead-lag time constants, and the gain. When the structure is more complex (e.g parallel SSDC control) there are naturally more parameters to optimize, hence, complexifying the tuning methodologies.

\subsubsection{SSDC design}

Ideally, the choice of the listed parameters should be done by eigenvalue placement and modal analysis \cite{c_karawita_c4-b4_2023}. However, in addition to the complexity and the non-linearities of the system, the idea is to keep supposing black-box models. This means that only empiric methods are to be adopted. Therefore, one option is to use the test signal method \cite{wang_sso_2017} to optimize the parameters. This is the same idea as the perturb-and-observe of the complex torque coefficients method, where a perturbation is injected, and then, the measure of variables allows tuning of phase shift.

In this paper, the adopted design process of the SSDC is:
\begin{itemize}
  \item The input signal of the SSDC is the frequency of the network, $\omega_{net}$, estimated by the PLL of the HVDC link.
  \item The band-pass filter is a second order filter with a quality factor of 50.
  \item The lead-lag parameters ($T_1$, $T_2$) are chosen using a method based on the optimisation of the electrical damping.
  \item The gain value is chosen at fixed phase shift and must provide a positive total damping in the worst situation.
  \item The limit of the output is set to $\pm 0.05pu$ or 5\% of the nominal active power.
  \item The output of the SSDC will be the active power of the HVDC link.
\end{itemize}

\begin{figure}[!htbp]
\centering
\includegraphics[width=0.99\columnwidth]{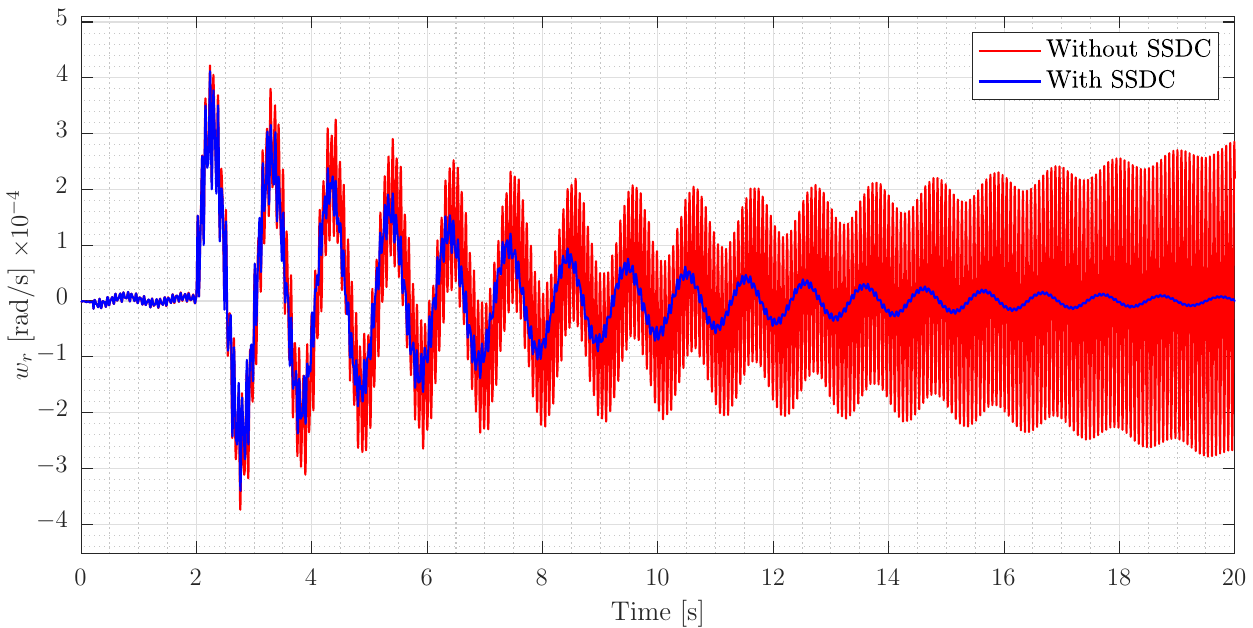}
\caption{Rotor speed deviation in an unstable situation with tuned SSDC and without SSDC.}
\label{wr_SSDC}
\end{figure}

For the use case, it was decided to only damp the first mode, as it is the only unstable mode. So, the tuning of the lead-lag parameters is done by computing iteratively the electrical damping at mode 1 frequency with a constant gain. The maximum damping phase shift $\Delta\Phi$ is kept, and the time constants are computed to center the transfer function on the damped frequency with: 
\begin{align}
 a = \frac{T_2}{T_1} = \frac{1-sin(\Delta\Phi)}{1+sin(\Delta\Phi)}\label{eq:a}
\end{align}
\begin{align}
 T_1 = \frac{1}{2\pi f_1a}\label{eq:T1}
\end{align}
To optimise the gain $K_{SSDC}$, it is also iterated on different values of gain, and the evolution of the damping is plotted. The gain value $K_{SSDC}$ should provide a maximal damping without reaching the controller output limits (see Fig.~\ref{SSDC}), since they induce harmonics that can feed negatively on the oscillations. With the correct tuning, the temporal results Fig.~\ref{wr_SSDC} shows that a well-tuned SSDC can damp SSO remotely (without communications) and effectively a power plant in the proximity. The impact of SSDC on the dynamic of the VSC under a Fault ride through (FRT) has been investigated and no changes were visible.

\begin{figure}[!htbp]
\centering
\includegraphics[width=0.99\columnwidth]{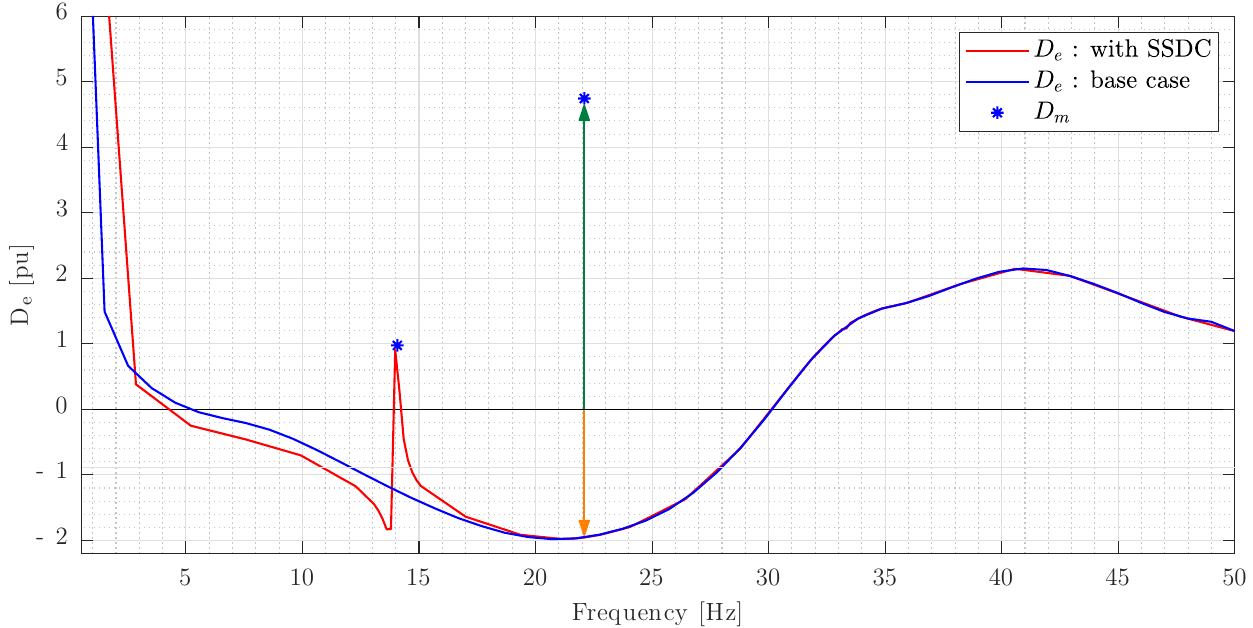}
\caption{Electrical damping in function of the frequency with (red) and without (blue) SSDC.}
\label{fig:De_SSDC}
\end{figure}

By performing the $D_e$ calculation of the system considering the SSDC, the results from Fig.~\ref{fig:De_SSDC} are obtained, where it is highlighted that the impact on the other frequencies of the subsynchronous range is negligible. Only the electrical at mode  frequency is increased. 

\subsection{SSTI protection Design}
\begin{figure}[!htbp]
\centering
\includegraphics[width=0.99\columnwidth]{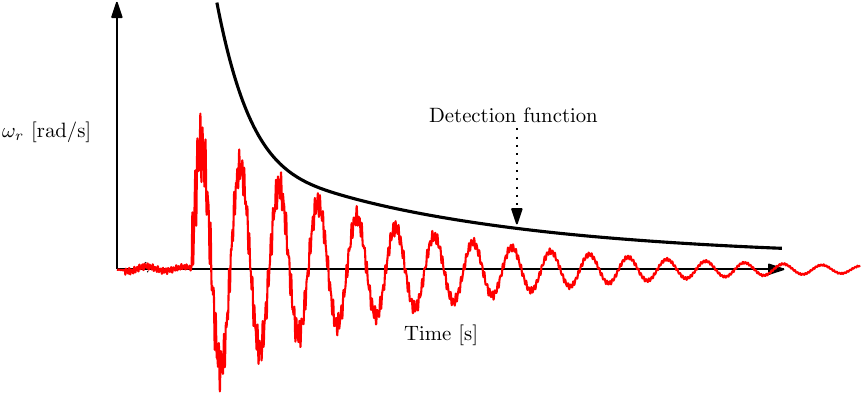}
\caption{Detection function with stable oscillations\cite{perera_protection_2018}.}
\label{fig:Detection_function}
\end{figure}

Another aspect of the SSTI risk is the protection of the power plants. The mitigation goal is to reduce the risk and to prevent most of the risky situations without impacting the electricity production. The protection objective is to ensure that the power plant will never be in a situation threatening its integrity. Thus, the protection has to detect the SSO, evaluate the risk and disconnect the power plant to stop the interactions if needed.
To do so, measures of current or voltage are done to evaluate the magnitude of the oscillations \cite{perera_protection_2018}. Then, with this estimation of the magnitude, a non-linear detection function is applied. Fig.~\ref{fig:Detection_function} represents a detection function. There is no trip if the oscillation after a perturbation remains below the detection curve: the system is stable. On the contrary, if the SSO takes too long to decrease, it means that the damping is not sufficient and that the shaft is at risk. The protection of the power plant will open and the production stop.

\section{Discussions}
\subsection{Limitations of the methodology}

This article uses different indicators to evaluate the stability of the system. First, the UIF gives an idea of the stability with a high-level model. Discussions are needed to determine a UIF threshold for VSC-HVDC. Then, with the complex torque coefficients method, we chose to evaluate both the electrical damping and the mechanical damping. Still, the mechanical damping is usually not evaluated as it can be difficult to estimate \cite{saad_study_2018}. It is neglected and the system is considered as stable when the electrical damping is positive. This allows a margin when evaluating and mitigating the risk.


\subsection{Data and resources}

To be able to perform the SSO derisking study, some information and resources need to be gathered. This supposes the collaboration of the different actors to provide data at each step of the study.

\subsubsection{UIF data}
For the screening part of the study, the reason UIF is widely used is because of its simplicity and because of the few information needed. Indeed, the detailed models of the different systems are not used for the screening study, only the basic electrical models. The minimal information required for the computation of this indicator can be found in Table~\ref{table_data_uif}.

\begin{table}[!htbp]
\renewcommand{\arraystretch}{1.3}
\setlength{\tabcolsep}{3pt}
\centering
\caption{UIF data.}
\label{table_data_uif}
\begin{tabular}{|c|c|c|c|}
\hline
Info. & Owner & Info. & Owner\\
\hline
$V_{AC}$ & \makecell{TSO \\ (Transport System Operator)}& $X_{eq,Gen}$ & \makecell{Power Plant \\ operator}\\
\hline
$S_{sc,Grid}$ & TSO & $S_{\text{HVDC}}$ & \makecell{HVDC link \\ operator}\\
\hline
$X_{lignes}$ & TSO & $X_{\text{eq,HVDC}}$ & \makecell{HVDC link \\ operator}\\
\hline
$S_{Gen}$ & \makecell{Power Plant \\ operator} & \makecell{Network \\ structure} & TSO \\
\hline
\end{tabular}
\end{table}

The computing resources to calculate UIF are negligible as it is a simple equation.

\subsubsection{Complex torque coefficients data}

For the computation of the damping (mechanical and electrical) more detailed information and models from the power plant operator and the HVDC link operator need to be shared. For the mechanical model, the power plant operator needs to provide the modal frequencies of the shaft and the corresponding mechanical damping (if used). Then, for the electrical damping a black-box single-mass and multi-mass model of the power plant, a manufacturer black-boxed model of the HVDC link and a model of the network are needed. The black-boxes must have configurable functioning point and the adequate input and output to perform the studies: injection of perturbations, measures... Data needs are summarised in Table~\ref{table_data_De}. The responsible of who shall perform the study of derisking needs to be defined. This is important because the models and the data are sensible information, and confidentiality has to be respected. Moreover, each actor has the responsibility of their model and assets; effort should be made to keep models representative and up to date.

\begin{table}[!htbp]
\renewcommand{\arraystretch}{1.3}
\centering
\caption{Complex torque coefficients method data.}
\label{table_data_De}
\begin{tabular}{|c|c|}
\hline
Parameter names & Owner\\
\hline
HTB network structure & TSO \\
\hline
$S_{sc,Grid}$ & TSO\\
\hline
$X_{lignes}$ & TSO\\
\hline
Black box power plant & Power Plant operator\\
\hline
Black box HVDC link & Power Plant operator\\
\hline
Modal frequencies of the shaft & Power Plant operator\\
\hline
Mechanical damping & Power Plant operator\\
\hline
Black box HVDC link & HVDC link operator\\
\hline
\end{tabular}
\end{table}

\subsubsection{SSDC data} 

To design an SSDC damping system the same data as for the complex torque coefficients method are needed. Indeed, the method used to develop the SSDC is based on the optimisation of the electrical damping. So, the computation of this damping at some frequencies is used.


\subsubsection{Challenges}

The gathering and communication of data between the actors is a challenge. As highlighted in \cite{c_karawita_c4-b4_2023}, the models can be under NDA, not compatible with the study provider software, badly documented, too old, or even may not exist in some cases. The quality of the models and their availability are something actors involved in SSO studies have to work with to make SSO studies more accurate and to guarantee stability of the future grid.

\section{Conclusion}

The methods to perform SSO derisking of an industrial project already exist. Even if some discussions about the application of some criteria are still needed, the main ideas are already applied for decades. Nevertheless, in this article it is proposed a clarification of some aspects by application of the methodology to a realistic study. It has been pointed out that many data from different actors are used in the SSO studies. The collaboration and share of information are important to achieve a complete and accurate derisking. However, developing and keeping models up to date is a real challenge for all the operators. This article emphasizes that the use of black box models may be sufficient to perform a full SSO derisking study but also to mitigate the risk using hardware or SSDC solutions. The computing resources are not a real problem while the models remain light but increasing computation time for heavier studies is to expect. So, optimisation of the algorithms and use of simplified models may be needed.

\section{Acknowledgement}

We thank Valentin Costan for his assistance, guidance and provision of initial simulation EMTP-RV models.

\bibliographystyle{IEEEtran}
\bibliography{PSCC}

\end{document}